# Interpretable Machine Learning for Genomics

David S. Watson[1]

[1]Department of Statistical Science, University College London, London, UK

Email for correspondence: david.watson@ucl.ac.uk

## §0   Abstract

High-throughput technologies such as next generation sequencing allow biologists to observe cell function with unprecedented resolution, but the resulting datasets are too large and complicated for humans to understand without the aid of advanced statistical methods. Machine learning (ML) algorithms, which are designed to automatically find patterns in data, are well suited to this task. Yet these models are often so complex as to be opaque, leaving researchers with few clues about underlying mechanisms. Interpretable machine learning (iML) is a burgeoning subdiscipline of computational statistics devoted to making the predictions of ML models more intelligible to end users. This article is a gentle and critical introduction to iML, with an emphasis on genomic applications. I define relevant concepts, motivate leading methodologies, and provide a simple typology of existing approaches. I survey recent examples of iML in genomics, demonstrating how such techniques are increasingly integrated into research workflows. I argue that iML solutions are required to realize the promise of precision medicine. However, several open challenges remain. I examine the limitations of current state of the art tools and propose a number of directions for future research. While the horizon for iML in genomics is wide and bright, continued progress requires close collaboration across disciplines.

## §1   Introduction

Technological innovations have made it relatively cheap and easy to observe biological organisms at molecular resolutions. High-throughput methods such as next generation sequencing and the full suite of "omic" platforms – e.g., genomic, proteomic, metabolomic, and related technologies – have inaugurated a new era of systems biology, providing data so abundant and detailed that researchers are not always sure just what to do with the newfound embarrassment of riches. One of the most salient traits of these datasets is their sheer size. Sequencing technologies can record anywhere from a few thousand to a few billion features per sample. Another important factor, related but distinct, is that genomic data is not immediately intelligible to humans. Whereas a small child can accurately classify pictures of animals, experts cannot generally survey a genetic sequence and predict health outcomes – at least not without the aid of advanced statistical models.

Machine learning (ML) algorithms are designed to automatically mine data for insights using few assumptions and lots of computational power. With their ability to detect and exploit complex relationships in massive datasets, ML techniques are uniquely suited to the challenges of modern genomics. In this article, I will focus specifically on *supervised* learning methods, which attempt to estimate a function from inputs (e.g., gene expression) to outputs (e.g., disease diagnosis).[1] ML algorithms have become enormously popular in medical research (Topol, 2019), especially in imaging tasks such as radiological screening (Mazurowski et al., 2019) and tumor identification (McKinney et al., 2020). They have also been successfully applied to complex molecular problems such as antibiotic discovery (Stokes et al., 2020) and predicting regulatory behavior from genetic variation (Eraslan et al., 2019). ML promises to advance our understanding of fundamental biology and revolutionize the practice of medicine, enabling personalized treatment regimes tailored to a patient's unique biomolecular profile.

Despite all their strengths and achievements, supervised learning techniques pose a number of challenges, some of which are especially troubling in biomedical contexts. Foremost among these is the issue of *interpretability*. Successful ML models are often so complex that no human could possibly follow the reasoning that leads to individual predictions. Inputs may pass through a long sequence of recursive nonlinearities, spanning thousands or millions of parameters, before a prediction emerges out the other side. How can such a black box, no matter how accurate, advance our knowledge of biological mechanisms? How can we trust what we do not understand?

Interpretable machine learning (iML) – also known as explainable artificial intelligence (xAI) or, more simply, explainability – is a fast-growing subfield of computational statistics devoted to helping users make sense of the predictions of ML models. Cataloging the state of the art in iML has become a whole meta-literature unto itself. Recent examples include (but are almost certainly not limited to): Adadi & Berrada (2018); Gilpin et al. (2018); Guidotti et al. (2018); Mueller et al. (2019); Murdoch et al. (2019); Barredo Arrieta et al. (2020); Das & Rad (2020); Mohseni et al. (2020); Vilone & Longo (2020); Xu et al. (2020); Marcinkevičs & Vogt (2020); Linardatos et al. (2021); and Rudin et al. (2021). Holzinger et al. (2019) frame their survey with a focus on medical applications, while Azodi et al. (2020) specialize more narrowly on iML for genetics. For reviews of interpretable deep learning in genomics, see Talukder et al. (2020) and Binder (this issue).

My goal in this article is not to add yet another survey to this overpopulated field. Instead, I aim to provide a gentle and critical introduction to iML for genomic researchers. I define relevant concepts, motivate prominent approaches, and taxonomize popular method-ologies. I examine important opportunities and challenges for iML in genomics, arguing that more intelligible algorithms will ultimately advance our understanding of systems biology and play a crucial role in realizing the promise of precision medicine. I show how iML algorithms augment the traditional bioinformatics workflow with explanations that can be used to guide data collection, refine training procedures, and monitor models throughout deployment (see Fig. 1). To do so, however, the field must overcome several conceptual and technical obstacles. I outline these issues and suggest a number of future research directions, all of which require close interdisciplinary collaboration.

---

[1] This is distinct from *unsupervised* learning, which searches for structure without predicting outcomes, and *reinforcement* learning, where agents select a policy to maximize rewards in a particular environment. Though both methods have been applied in genomics, supervised learning is more prevalent and remains the focus of almost all contemporary research in interpretability.

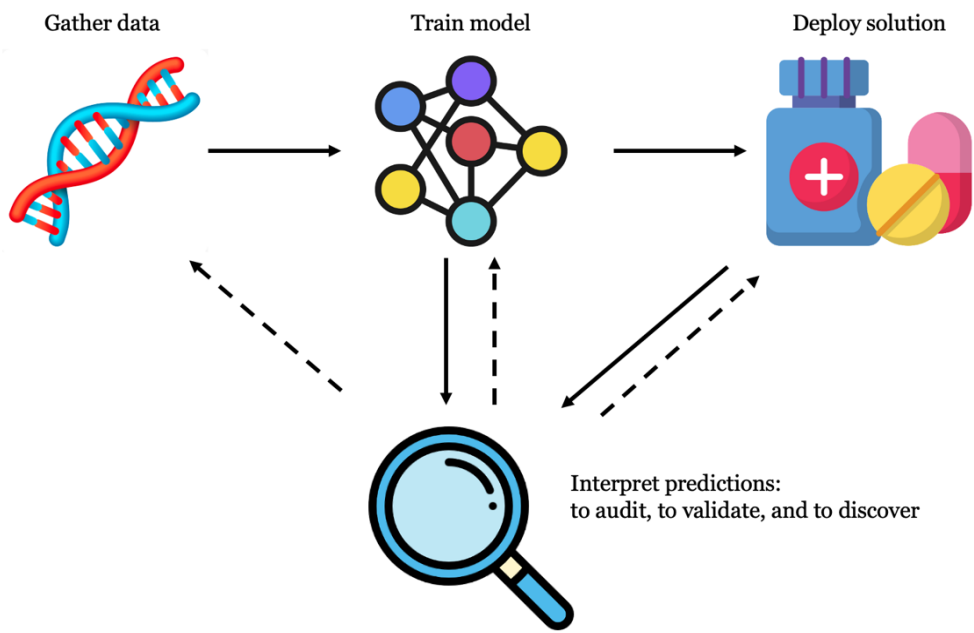

**Figure 1**. The classic bioinformatics workflow spans data collection, model training, and deployment. iML augments this pipeline with an extra interpretation step, which can be used during training and throughout deployment (incoming solid edges). Algorithmic explanations (outgoing dashed edges) can be used to guide new data collection, refine training, and monitor models during deployment.

The remainder of this article is structured as follows. I review relevant background material and provide a typology of iML in Sect. 2. I examine three motivations for iML in Sect. 3, each of which has a role to play in computational biology. In Sect. 4, I introduce a number of popular iML methodologies that have been used in recent genomic research. A discussion follows in Sect. 5, where I consider three open challenges for iML that are especially urgent in bioinformatics. Sect. 6 concludes.

## §2  Background

In supervised learning, we assume access to a finite training dataset of $n$ input-output pairs, $\{x_i, y_i\}_{i=1}^{n}$. The vector $x_i$ denotes a point in $p$-dimensional space, $x_i = (x_{i1}, \ldots, x_{ip})$, with coordinate $x_{ij}$ corresponding to the $i^{\text{th}}$ value of feature $X_j$. For instance, $x_i$ may represent an individual's transcriptomic profile and $x_{ij}$ their read count for gene $X_j$. Samples are presumed to be independently and identically distributed instances of some fixed but unknown joint distribution $P(X, Y)$. The goal is to infer a function $f: \mathcal{X} \to \mathcal{Y}$ that maps *x*-vectors to *y*-outcomes. For example, we may want to group cancer patients into subtypes with different prognostic trajectories based on gene expression. If outcomes are continuous, we say that $f$ is a regressor; if they are categorical, we say that $f$ is a classifier. In either case, performance is evaluated via some loss function that quantifies model error. While the expected loss – also known as the *risk* – cannot be directly calculated without knowledge of the underlying distribution, it can be estimated either on the training data or, preferably, on an independent test set sampled from the same distribution. Empirical risk minimization is the learning strategy whereby we select the top performing model from some predefined function class (Vapnik, 1995). Popular algorithms of this type include (potentially regularized) linear models, neural networks, and tree-based ensembles such as random forests and gradient boosting machines. See (Hastie, Tibshirani, & Friedman, 2009) for a good introduction.

It is not always obvious what would constitute a successful explanation of a given model prediction. Indeed, explanation itself is an epistemologically contested concept, the subject of ancient and modern philosophical debates (Woodward, 2019). It should perhaps come as no surprise then to learn that algorithmic explanations come in many flavors. The first point to acknowledge is that iML tools are used for different analytic purposes. For instance, they may help to estimate or understand a true functional relationship presumed to hold in nature. Alternatively, they may be used to analyze the behavior of a fitted model – to illuminate the black box, as it were. Finally, they may be involved in the design of so-called "glass box" algorithms, i.e. some novel function class specifically built for transparency. These goals may overlap at the edges, and methods originally intended for one may be repurposed for another. However, each represents a distinct task with its own challenges. Not all are equally prevalent in genomics, though this review will discuss examples of each. Keeping these aims separate is crucial to avoid the conceptual pitfalls addressed in Sect. 5.

The following typology is adapted from Molnar's (2021) textbook guide to iML, which provides a helpful overview of technical approaches and the current state of the art. Roughly put, there are three key dichotomies that orient iML research: intrinsic vs. post-hoc, model-specific vs. model-agnostic, and global vs. local. A final consideration is what type of output the method generates. Each point is considered in turn.

*Intrinsic vs. post-hoc.* An intrinsically explainable algorithm is one that raises no intelligibility issues in the first place – i.e., a glass box algorithm. Canonical examples include (sparse) linear regression and (short) rule lists. Parametric models typically include interpretable parameters that correspond to meaningful quantities, e.g. a linear coefficient denoting a gene's log fold change in a microarray experiment. Some have argued that such models are the only ones that should be allowed in high-risk settings, such as those arising in healthcare (Rudin, 2019). Unfortunately, many interesting real-world problems cannot be adequately modelled with intrinsically explainable algorithms. Watson et al. (2019) argue that in clinical medicine, doctors are obligated to use whatever available technology leads to the best health outcomes on average, even if that involves opaque ML algorithms. Of course, flexible ML models are also prone to overfit the training data, especially in high-dimensional settings. The choice of which approach to use invariably depends on contextual factors – the task at hand, the prior knowledge available, and what assumptions the analyst deems reasonable. Should researchers choose to use some black box method, interpreting predictions will require post-hoc tools, which take a target model $f$ as input and attempt to explain its predictions, at least near some region of interest.

*Model-specific vs. model-agnostic.* Model-specific iML solutions take advantage of the assumptions and architectures upon which particular algorithms are built to generate fast and accurate explanations. For example, much work in iML has been specifically devoted to deep neural networks (Bach et al., 2015; Montavon et al., 2017; Shrikumar, Greenside, & Kundaje, 2017; Sundararajan, Taly, & Yan, 2017), an especially rich class of functions with unique explanatory affordances and constraints. Model-agnostic tools, on the other hand, strive for more general applicability. Treating the fitted function $f$ as a black box, they attempt to explain its predictions with few or no assumptions about the data generating process. Model-agnostic approaches are especially useful in cases where $f$ is inaccessible (for example if an algorithm is protected by intellectual property laws), while model-specific methods are generally more efficient and reliable when $f$'s structure is known.

*Global vs. local.* A global explanation helps the user understand the behavior of the target model $f$ across all regions of the feature space. This is difficult to achieve when $f$ is complex and/or high-dimensional. A local explanation, by contrast, is only meant to apply to the area near some particular point of interest. For instance, a properly specified linear regression is globally explainable in the sense that the model formula holds with equal probability for any randomly selected datapoint. However, a local linear approximation to some nonlinear $f$ will fit best near the target point, and does not in general tell us anything about how the model behaves in remote regions of the feature space. In biological contexts, we can think of global and local explanations applying at population- and individual-levels, respectively. These are poles of a spectrum that also admits of intermediate alternatives, e.g. subpopulation- or group-level explanations, which are possible as well.

A final axis of variation for iML tools is their output class. Typically, these methods explain predictions through some combination of images, statistics, and/or examples. Visual explanations are especially well suited to image classifiers (see Fig. 2). Other common visual approaches include plots that illustrate the partial dependence (Friedman, 2001) or individual conditional expectation (Casalicchio, Molnar, & Bischl, 2019) of variables, which can inform users about feature interactions and/or causal effects (Zhao & Hastie, 2019). Statistical outputs, by contrast, may include rule lists, tables, or numbers quantifying explanatory value in some predefined way. Finally, exemplary methods report informative datapoints, either in the form of prototypes, which typify a given class (Chen et al., 2019), or counterfactuals, which represent the most similar sample on the opposite side of a decision boundary (Wachter et al., 2018). These latter methods are less common in genomics, although conceptually similar matching algorithms are used in clinical medicine (Bica et al., 2021).

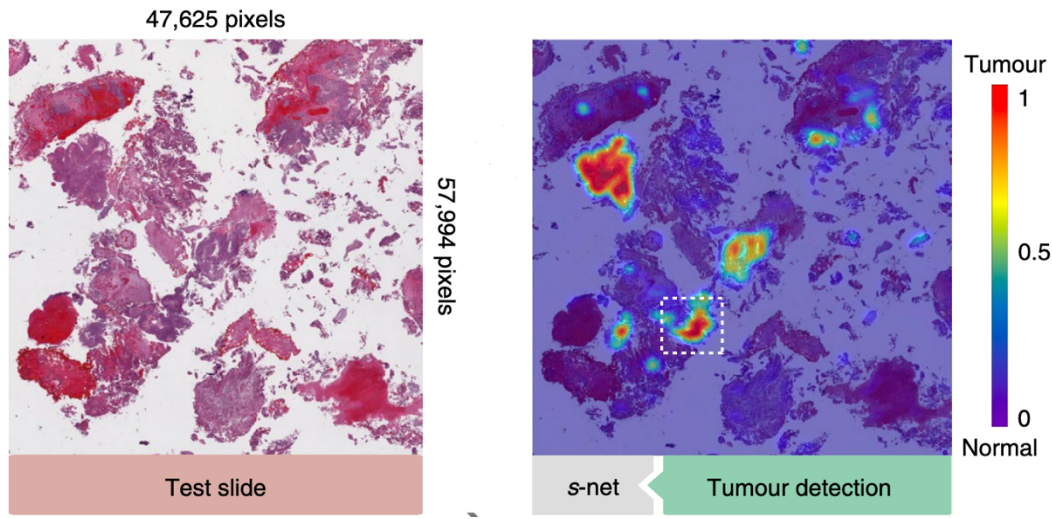

**Figure 2**. A saliency map visually explains a cancer diagnosis based on whole-slide pathology data. The highlighted regions on the right pick out the elements of the image that the algorithm deemed most strongly associated with malignancy. From (Zhang et al., 2019, p. 237).

## §3 Motivations

Why do we seek algorithmic explanations in the first place? Watson & Floridi (2020) offer three reasons: (1) to audit for potential bias; (2) to validate performance, guarding against unexpected errors; and (3) to discover underlying mechanisms of the data generating process. All three are relevant for genomics.

### §3.1 To Audit

Healthcare often magnifies social inequalities. For recent evidence, look no further than the COVID-19 pandemic, which disproportionately affects minority populations in the US and UK (Egede & Walker, 2020). ML threatens to automate these injustices. Obermeyer et al. (2019) found evidence of significant racial bias in a healthcare screening algorithm used by millions of Americans. Simulations suggest that rectifying the disparity would nearly triple the number of Black patients receiving medical care. Similar problems are evident in genomic research. Individuals of white European ancestry make up some 16% of the global population, but constitute nearly 80% of all genome-wide association study (GWAS) subjects, raising legitimate concerns that polygenic risk scores and other tools of precision medicine may increase health disparities (Martin et al., 2019). As genomic screening becomes more prevalent, there will be substantial and justified pressure to ensure that new technologies do not reinforce existing inequalities. Algorithmic fairness and explainability may even be legally required under the European Union's 2018 General Data Protection Regulation (GDPR), depending on one's interpretation of the relevant articles (Selbst & Powles, 2017; Wachter et al., 2017). By making a model's reliance on potentially sensitive attributes more transparent, iML methods can help quantify and mitigate potential biases.

### §3.2 To Validate

The second motivation concerns the generalizability of ML models. Supervised learning algorithms are prone to overfitting, which occurs when associations in the training data do not generalize to test environments. In a famous example, a neural network trained on data from a large New York hospital classified asthmatics with pneumonia as low risk, a result that came as a surprise to the doctors and data scientists working on the project (Caruana et al., 2015). The algorithm had not uncovered some subtle pulmonological secret. On the contrary, the apparent association was spurious. Asthmatics with pneumonia are at such great risk that emergency room doctors immediately send them to the intensive care unit, where their chances of survival are relatively high. It was the extra medical attention, not the underlying condition, that improved outcomes for these patients. Naively applying this algorithm in a new environment – e.g., some hospital where patient triage is performed by a neural network with little or no input from doctors – could have grave consequences.

Overfitting has been observed in GWAS models (Nicholls et al., 2020), where associations that appear informative in one population do not transfer to another. Such failures can be difficult to detect given the complexity of the underlying signals, which may depend on environmental factors or subtle multi-omic interactions. The problem of *external validity* or *transportability* is well known in the natural and social sciences, if not always well understood (Bareinboim & Pearl, 2016; Pearl & Bareinboim, 2014). Because causal dependencies are robust to perturbations of upstream variables, environmental heterogeneity – e.g., different patient subpopulations or data collection protocols – can help isolate and quantify causal effects (Heinze-Deml et al., 2018; Peters et al., 2016). These methods have motivated novel GWAS methodologies that search for persistent associations across varying patterns of linkage disequilibrium (Li et al., 2021). iML algorithms, together with causal inference tools (Imbens & Rubin, 2015; Pearl, 2000; Peters et al., 2017), can help researchers identify and remove spurious signals, ensuring better generalizability to new environments.

### §3.3 To Discover

A final goal of iML, less widely discussed than the previous two but arguably of greater interest in genomics, is to reveal unknown properties and mechanisms of the data generating process. In this case, the guiding assumption is not that the target model $f$ is biased or overfit; on the contrary, we assume that it has found some true signal and investigate its internal logic to learn more. This may mean examining weights from a support vector machine, approximating a decision boundary with some local linear model, or extracting Boolean rules to describe the geometry of some complex regression surface. Examples of all these approaches and more will be examined below, demonstrating how iML can be – and to some extent already has been – integrated into genomic research workflows. By unpacking the reasoning that underlies high-performance statistical models, iML algorithms can mine for insights and suggest novel hypotheses in a flexible, data-driven manner. Rightly or wrongly, it is this capacity – not its potential utility for auditing or validation – that is likely to inspire more widespread adoption in bioinformatics.

### §4 Methodologies and Applications

In this section, I introduce a number of prominent approaches to iML. As a running example, I will consider a hypothetical algorithm that classifies breast cancer patients into different subtypes on the basis of a diagnostic biomarker panel. Such tools have been in use for decades, although recent advances in ML have vastly increased their accuracy and sophistication (Cascianelli et al., 2020; Sarkar et al., 2021). I examine three particular iML methods at length – variable importance, local linear approximators, and rule lists – describing their basic forms and reviewing some recent applications. There is considerable variety within each subclass, and the choice of which to use for a given task is inevitably context dependent. While all three could be fruitfully applied for auditing, validation, or discovery, the latter has tended to dominate in genomic research to date.

### §4.1 Variable Importance

Variable importance (VI) measures are hardly new. Laplace and Gauss, writing independently in the early 19[th] century, each described how standardized linear coefficients can be interpreted as the average change in response per unit increase of a feature, with all remaining covariates held fixed. Coefficients with larger absolute values therefore suggest greater VI. If our cancer subtyping algorithm were linear, then we might expect large weights on genes such as BRCA1, which is strongly associated with basal-like breast cancer (BLBC) (Turner & Reis-Filho, 2006), and ESR1, a known marker of the luminal A subtype (Sørlie et al., 2003).

The ease of computing VI for linear models has been exploited to search for causal variants in single nucleotide polymorphism (SNP) arrays via penalized regression techniques like the lasso (Tibshirani, 1996) and elastic net (Zou & Hastie, 2005), both popular in GWAS (Waldmann et al., 2013). Random forests (Breiman, 2001a), one of the most common supervised learning methods in genomics (Chen & Ishwaran, 2012), can provide a range of marginal or conditional VI scores, typically based either on permutations or impurity metrics (Altmann et al., 2010; Nembrini et al., 2018; Strobl et al., 2008). Support vector machines (SVMs) may give intelligible feature weights depending on the underlying kernel (Schölkopf et al., 2004). For example, Sonnenburg et al. (2008) use a string kernel to predict splice sites in *C. elegans* and extract relevant biological motifs from the resulting positional oligomer importance matrices. The method has since been extended to longer sequence motifs and more general learning procedures (Vidovic et al., 2015; Vidovic et al., 2017). More recently, Kavvas et al. (2020) used an intrinsically interpretable SVM to identify genetic determinants of antimicrobial resistance (AMR) from whole genome sequencing data. Whether these methods tell us about the importance of features in nature or just in some fitted model $f$ depends on whether we assume that $f$ accurately captures the functional form of the relationship between predictors and outcomes.

To go back to the typology above, VI measures are global parameters that may be intrinsic (as in linear models) or post-hoc (as in random forest permutation importance). Model-specific versions are popular, although several model-agnostic variants have emerged in recent years. These include targeted maximum likelihood measures (Hubbard, Kennedy, & van der Laan, 2018; Williamson et al., 2020), nested model tests using conformal inference (Lei et al., 2018; Rinaldo et al., 2019), and permutation-based reliance statistics (Fisher et al., 2019). Such methods typically involve computationally intensive procedures such as bootstrapping, permutations, and/or model refitting, which pose both computational and statistical challenges when applied in settings with large sample sizes and/or high-dimensional feature spaces, such as those commonly found in genomics.

A notable exception specifically designed for high-dimensional problems is the knockoff test for variable selection (Barber & Candès, 2015; Candès et al., 2018). The basic idea behind this approach is to generate a set of "control" variables – the eponymous knockoffs – against which to test the importance of the original input features. For a given $n \times p$ design matrix $\boldsymbol{X}$, we say that $\widetilde{\boldsymbol{X}}$ is the corresponding knockoff matrix if it meets the following two criteria:

(a) *Pairwise exchangeability*. For any proper subset $S \subset [p] = (1, \dots, p)$:
$$(\boldsymbol{X}, \widetilde{\boldsymbol{X}})_{swap(S)} =^d (\boldsymbol{X}, \widetilde{\boldsymbol{X}}),$$
where $=^d$ represents equality in distribution and the swapping operation is defined below.

(b) *Conditional independence*. $\widetilde{\boldsymbol{X}} \perp Y \mid \boldsymbol{X}$.

A swap is obtained by switching the entries $X_j$ and $\tilde{X}_j$ for each $j \in S$. For example, with $p = 3$ and $S = \{1, 3\}$:

$$(X_1, X_2, X_3, \tilde{X}_1, \tilde{X}_2, \tilde{X}_3)_{swap(S)} =^d (\tilde{X}_1, X_2, \tilde{X}_3, X_1, \tilde{X}_2, X_3).$$

A supervised learning algorithm is then trained on the expanded $n \times 2p$ feature matrix, including original and knockoff variables. If $X_j$ does not significantly outperform its knockoff $\tilde{X}_j$ by some model-specific importance measure – Candès et al. (2018) describe methods for lasso and random forests – then the original feature may be safely removed from the final model. The authors outline an adaptive thresholding procedure that provably provides finite sample false discovery rate (FDR) control for variable selection under minimal assumptions. They also describe a conditional randomization test (CRT) for asymptotic type I error rate control, in which observed values are compared to null statistics repeatedly sampled from the knockoff distribution. Experiments suggest the CRT is more powerful than the adaptive procedure, and therefore could be preferable when signals are sparse. However, Candès et al. caution that the CRT may be infeasible for large datasets.

The most challenging aspect of this pipeline is computing the knockoff variables themselves. However, when good generative models are available, the technique can be quite powerful. Watson & Wright (2021) use Candès et al.'s original semidefinite programming formulation to approximate knockoffs for a DNA microarray experiment. Sesia et al. (2019) extend the method to genotype data using a hidden Markov model to sample knockoffs. In a recent study, Bates et al. (2020) used knockoffs with GWAS data from parents and offspring to create a "digital twin test" based on the CRT, in which causal variants are identified by exploiting the natural randomness induced by meiosis. The method has greater power and localization than the popular transmission disequilibrium test.

It should be noted that VI measures can be applied either to raw or processed features. The latter can be especially valuable when the original data is difficult to interpret. In a recent paper, Chia et al. (2020) use direct wavelet transform to capture the location and frequency of bacterial Raman spectra as a preprocessing step toward building a more interpretable classifier. They use knockoffs for (processed) feature selection and train a multinomial logistic regression, reporting performance on par with the best black box models. Of course, not all genomic settings have comparably interpretable low-dimensional representations. I revisit the issue of variable abstraction and granularity as an open challenge in Sect. 5.3.

## §4.2 Local Linear Approximators

All methods described above are global in scope. The goal in many iML settings, however, is to provide explanations for *individual predictions*. This could be useful if, for instance, a subject receives an unexpected diagnosis. Perhaps Alice shows few obvious signs of BLBC and deviates markedly from the classic patient profile, yet our algorithm assigns her to this class with high probability. Given the aggressive treatment regime likely to follow such a diagnosis, Alice wants to be certain that the classification is correct. A local explanation reveals that, though BRCA1 mutations account for many BLBC predictions, in her case, the feature is relatively unimportant. Instead, her local explanation turns largely on CXCR6 – a gene associated with the Basal I subtype, which has a better prognosis on average than Basal II, and is therefore less likely to require high doses of chemotherapy (Milioli et al., 2017).

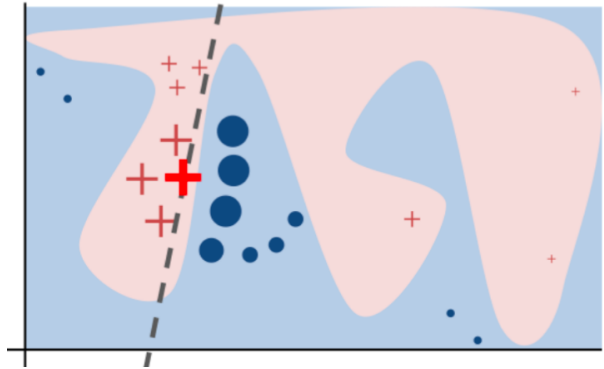

**Figure 3**. A complex decision boundary (the pink blob/blue background) separates red crosses from blue circles. This function cannot be well-approximated by a linear model, but the boundary near the large red cross is roughly linear, as indicated by the dashed line. From (Ribeiro et al., 2016, p. 1138).

Local linear approximators have become the de facto standard method for computing local explanations of this sort (Bhatt et al., 2020). These algorithms assign weights to each input feature that sum to the model output. The idea derives from the insight that even though target functions may be highly complex and nonlinear, any point on a continuous curve will have a linear tangent (see Fig. 3). By estimating the formula for this line, we can approximate the regression surface or decision boundary at a particular point. Popular examples of local linear approximators include LIME (Ribeiro et al., 2016) and SHAP (Lundberg & Lee, 2017), both of which are implemented in user-friendly Python libraries. The latter has additionally been incorporated into iML toolkits distributed by major tech firms such as Microsoft,[2] Google,[3] and IBM.[4] I will briefly explicate the theory behind this method, which unifies a number of similar approaches, including LIME.

SHAP is founded on principles from cooperative game theory, where Shapley values were originally proposed as a way to fairly distribute surplus across a coalition of players (Shapley, 1953). In iML settings, players are replaced by input features and Shapley values measure their contribution to a given prediction. Let $x_i \in \mathbb{R}^p$ denote an input datapoint and $f(\pmb{x}_i) \in \mathbb{R}$ the corresponding output of function $f$. Shapley values decompose this number into a sum of feature attributions:

$$f(\pmb{x}_i) = \sum_{j=0}^{p} \phi_j,$$

where $\phi_0$ denotes a baseline expectation (e.g., the mean response) and $\phi_j$ ($j \geq 1$) the weight assigned to feature $X_j$ at point $\pmb{x}_i$. Let $v: 2^p \to \mathbb{R}$ be a value function such that $v(S)$ is the payoff associated with feature subset $S \subseteq [p]$ and $v(\{\emptyset\}) = 0$. The Shapley value $\phi_j$ is given by $j$'s average marginal contribution to all subsets that exclude it:

$$\phi_j = \frac{1}{p!} \sum_{S \subseteq [p] \setminus \{j\}} |S|!\,(p - |S| - 1)!\,[v(S \cup \{j\}) - v(S)].$$

---

[2] See https://github.com/interpretml/interpret.
[3] See https://pair-code.github.io/what-if-tool/.
[4] See http://aix360.mybluemix.net/.

It can be shown that this is the unique value satisfying a number of desirable properties, including efficiency, linearity, sensitivity, and symmetry.[5] Computing exact Shapley values is NP-hard, although several efficient approximations have been proposed (Sundararajan & Najmi, 2019).

There is some ambiguity as to how one should calculate payoffs on a proper subset of features, since $f$ requires $p$-dimensional input. Let $S$ and $R$ be a partition of $[p]$, such that we can rewrite any $\pmb{x}_i$ as a pair of subvectors $(\pmb{x}_i^S, \pmb{x}_i^R)$. Then the payoff for feature subset $S$ takes the form of an expectation, with $\pmb{x}_i^S$ held fixed while $\pmb{X}^R$ varies. Following Merrick & Taly (2020), we consider a general formulation of the value function, indexed by a distribution $\mathcal{D}_R$:

$$v_{\mathcal{D}_R}(S) = \mathbb{E}_{\pmb{X}^R \sim \mathcal{D}_R}[f(\pmb{x}_i^S, \pmb{X}^R)].$$

Popular options for $\mathcal{D}_R$ include the marginal distribution $P(\pmb{X}^R)$, which is the default choice in SHAP; the conditional $P(\pmb{X}^R | \pmb{x}_i^S)$, implemented in the R package `shapr` (Aas, Jullum, & Løland, 2019); and the interventional $P(\pmb{X}^R | do(\pmb{x}_i^S))$, recently proposed by Heskes et al. (2020). Each reference distribution offers certain advantages and disadvantages, but the choice of which to use is ultimately dependent upon one's analytical goals (see Sect. 5).

SHAP has been used to identify biomarkers in a number of genomic studies. A model-specific variant known as DeepSHAP – with close ties to related methods DeepLIFT (Shrikumar et al., 2017) and integrated gradients (Sundararajan et al., 2017), all techniques for explaining the predictions of deep neural networks – was recently used in conjunction with a model for predicting differential expression based on genome-wide binding sites on RNAs and promoters (Tasaki et al., 2020). The same tool was used to identify CpG loci in a DNA methylation experiment that best predicted a range of biological and clinical variables, including cell type, age, and smoking status (Levy et al., 2020). Yap et al. (2021) trained a convolutional neural network to classify tissue types using transcriptomic data, prioritizing top genes as selected by DeepSHAP. Another SHAP variant – TreeExplainer (Lundberg et al., 2020), which is optimized for tree-based ensembles such as random forests – was used to identify taxa in the skin microbiome most closely associated with various phenotypic traits (Carrieri et al., 2021). SHAP is also gaining popularity in mass spectrometry, where data heterogeneity can complicate more classical inference procedures (Tideman et al., 2020; Xie et al., 2020).

---

[5] For formal statements of the Shapley axioms, see Lundberg & Lee (2017).

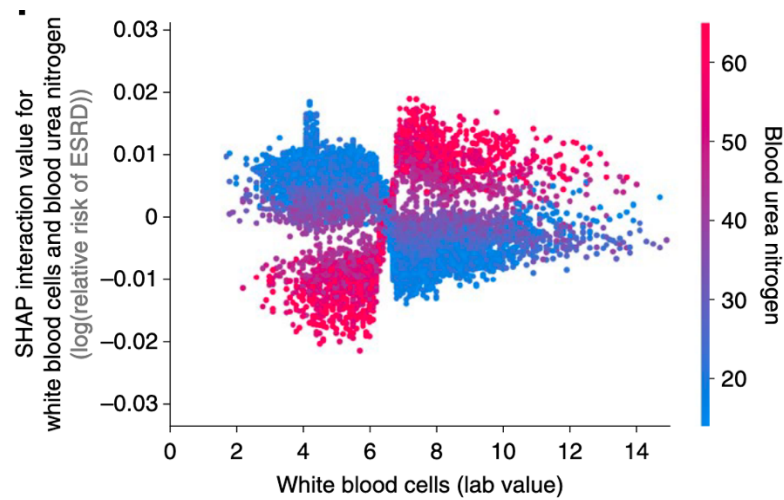

**Figure 4**. Shapley values show that high white blood cell counts increase the negative risk conferred by high blood urea nitrogen for progression to end stage renal disease (ESRD). From (Lundberg et al., 2020, p. 61).

In each of these cases, further investigation was required to confirm the involvement of selected features in the target functions. The outputs of SHAP, or any other iML algorithm for that matter, are by no means decisive or infallible. However, they offer a principled and novel approach for feature ranking and selection, as well as exploring interactions that can reveal unexpected mechanisms and guide future experiments. For instance, Lundberg et al. (2020) use Shapley interaction values to demonstrate that white blood cells are positively associated with risk of end stage renal disease (ESRD) in patients with high blood urea nitrogen, but negatively associated with ESRD in patients with low blood urea nitrogen (see Fig. 4). The multivariate nature of these attributions makes them more informative than the probe-level analyses common in differential expression testing, even when shrinkage estimators are used to "pool information" across genes (Law et al., 2014; Love, Huber, & Anders, 2014; Smyth, 2004). They are potentially more meaningful to individuals than global estimates such as those provided by knockoffs, since Shapley values are localized to explain particular predictions rather than average behavior throughout an entire feature space. With their model-agnostic flexibility and axiomatic underpinnings, local linear approximators are an attractive tool for genomic research.

## §4.3  Rule Lists

Rule lists are sequences of if-then statements, often visualized as a decision tree. Psychological studies have shown that humans can quickly comprehend explanations with such structures, at least when there are relatively few conditions, which is why rule lists are widely promoted as "intrinsically interpretable" (Lage et al., 2018). This accords with the privileged position of material implication in propositional logic, where → is typically regarded as a primitive relation, along with conjunction (∧), disjunction (∨), and negation (¬). These logical connectives form a functionally complete class, capable of expressing all possible Boolean operations. This flexibility, which allows for nonmonotonic and discontinuous decision boundaries, affords greater expressive power than linear models. Thus, if the true reason for Alice's unexpected diagnosis lies neither in her BRCA1 allele nor her CXCR6 expression but rather in some nonlinear interaction between the two, then she may be better off with a rule list that can concisely explain the (local or global) behavior of that function.

In statistical contexts, rule lists are generally learned through some process of recursive partitioning. For instance, the pioneering classification and regression tree (CART) algorithm (Breiman et al., 1984) predicts outcomes by dividing the feature space into hyperrectangles that minimize predictive error. Computing optimal decision trees is NP-complete (Hyafil & Rivest, 1976), but CART uses greedy heuristics that generally work well in practice. Because individual decision trees can be unstable predictors, they are often combined through ensemble methods such as bagging (Breiman, 2001a), in which predictions are averaged across trees trained on random bootstrap samples, and boosting (Friedman, 2001), in which predictions are summed over a series of trees, each sequentially optimized to improve upon the last. While combining basis functions tends to improve predictions, it unfortunately makes it difficult if not impossible to extract individual rules for better model interpretation. However, some regularization schemes have been developed to post-process complex learning forests for precisely this purpose. For instance, Friedman & Popescu (2008) propose the RuleFit algorithm, which mines a collection of Boolean variables by extracting splits from a gradient boosted forest. These engineered features are then combined with the original predictors in a lasso regression, producing a sparse linear combination of splits and inputs. Nalenz & Villani (2018) develop a similar procedure using a Bayesian horseshoe prior (Carvalho et al., 2010) instead of an $L_1$ penalty to induce shrinkage. They also add splits extracted from a random forest with those learned via gradient boosting to promote greater diversity.

Another strand of research in this area has focused on *falling* rule lists, which create monotonically ordered decision trees such that the probability of the binary outcome $Y = 1$ strictly decreases as one moves down the list. These models were originally designed for medical contexts, where doctors must evaluate patients quickly and accurately. For instance,

Letham et al. (2015) design a Bayesian rule list to predict stroke risk, resulting in a model that outperforms leading clinical diagnostic methods while being small enough to fit on an index card. Falling rule lists can be challenging to compute – see the note above about NP-completeness – and subsequent work has largely focused on efficient optimization strategies. Specifically, researchers have developed fast branch-and-bound techniques to prune the search space and reduce training time (Chen & Rudin, 2018; Yang et al., 2017), culminating in several tree-learning methods that are provably optimal under some restrictions on the input data (Angelino et al., 2018; Hu et al., 2019).

Less work has been done on localized rule lists, but there have been some recent advances in this direction. Ribeiro et al. (2018) followed up on their 2016 LIME paper with a new method, Anchors, which combines graph search with a multi-armed bandit procedure to find a minimal set of sufficient conditions for a given model prediction. Guidotti et al. (2018) introduce LORE, which simulates a balanced dataset of cases using a genetic algorithm designed to sample heavily from points near the decision boundary. A decision tree is then fit to the synthetic dataset. Lakkaraju et al. (2019)'s MUSE algorithm allows users to specify particular features of interest. Explanations are computed as compact decision sets within this subspace.

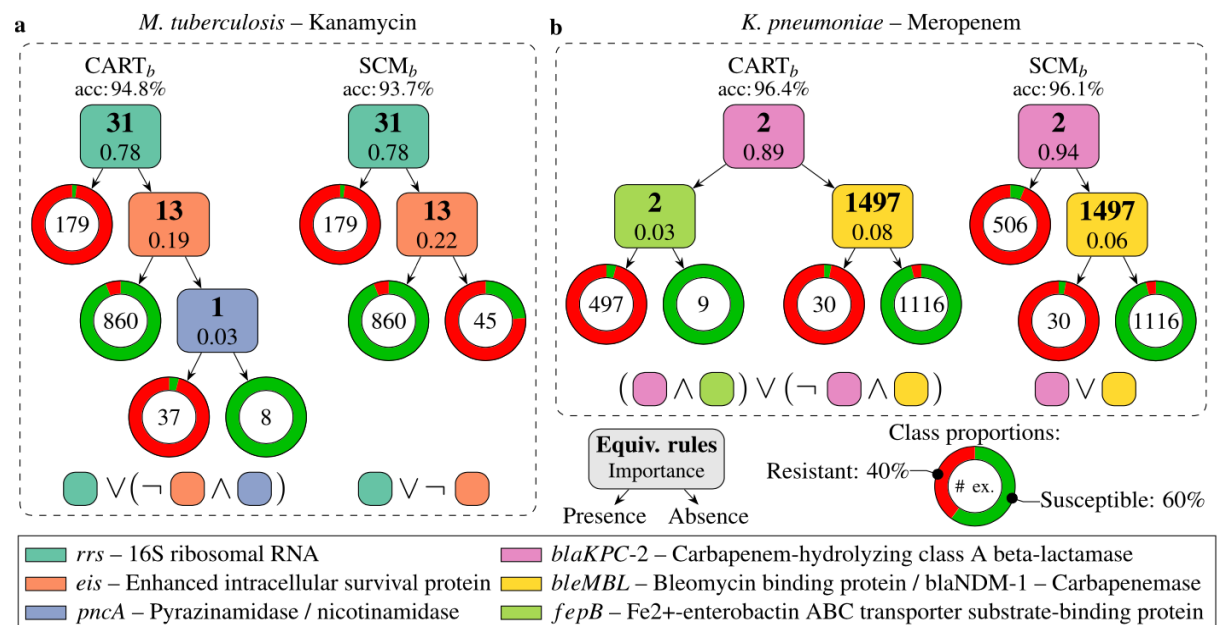

**Figure 5**. Example rule lists for AMR prediction from genotype data. Each rule detects the presence/absence of a *k*-mer and is colored according to the genomic locus at which it was found. From (Drouin et al., 2019, p. 4).

To date, rule lists have not been as widely used in genomics as feature attributions or local linear approximations. This likely has more to do with computational obstacles than any preference for particular model assumptions, per se. Still, some recent counterexamples buck the trend. Drouin et al. (2019) combine sample compression theory with recursive partitioning to learn interpretable genotype-to-phenotype classifiers with performance guarantees. As depicted in Fig. 5, these lists – visualized as trees and formulated as logical propositions below – can predict AMR in *M. tuberculosis* and *K. pneumoniae* with high accuracy using just a small handful of indicator functions over the space of all *k*-mers. Though their experiments focus on AMR, the method can be applied more generally. Anguita-Ruiz et al. (2020) use a sequential rule mining procedure to uncover gene expression patterns in obese subjects from longitudinal DNA microarray data. Garvin et al. (2020) combined iterative random forests (Cliff et al., 2019), a method for gene regulatory network inference, with random intersection trees (Shah & Meinshausen, 2014), which detect stable interactions in tree-based ensembles, to discover potentially adaptive SARS-CoV-2 mutations.

## §5   Open Challenges

Despite all the recent progress in iML, the field is still struggling with several challenges that are especially important in genomics. I highlight three in particular: ambiguous targets, error rate control, and variable granularity.

### §5.1   Ambiguous Targets

I have done my best above to be clear about the distinction between two tasks for which iML is often used: to better explain or understand (a) some fitted model $f$, or (b) some natural system that $f$ models. It is not always obvious which goal researchers have in mind, yet model- and system-level analyses require entirely different tools and assumptions. Whereas a supervised learning algorithm does not generally distinguish between correlation and causation, the difference is crucial in nature. Clouds predict rain and rain predicts clouds, but the causal arrow runs in only one direction. Genomic researchers face a fundamental ambiguity when seeking to explain, say, why Alice received her unexpected algorithmic diagnosis. Is the goal to explain why the model made the prediction it did, independent of the ground truth? Or, alternatively, is the goal to understand what biological conditions led to the diagnosis? The former, which I will call a *model-level* explanation, may be preferable in cases of auditing or validation, where the analyst seeks merely to understand what the algorithm has learned, without any further restrictions. In this case, we do not necessarily assume that the model is correct. The latter, which I will call a *system-level* explanation, is more useful in cases of discovery and/or planning, where real-world mechanisms cannot be ignored. In such instances, we (tentatively) presume that the prediction in question is accurate, at least to a first approximation.

   Model-level explanations are generally easier to compute, since features can be independently perturbed one at a time. This is the default setting for popular iML tools such as LIME and SHAP. System-level explanations, by contrast, require some structural assumptions about dependencies between variables. Such assumptions may be difficult or even impossible to test, raising legitimate questions about identifiability and underdetermination. Yet, as Pearl (2000) has long argued, there is value in articulating one's assumptions clearly, opening them up for scrutiny and debate instead of burying them behind defaults. The last year has seen a burst of new papers on causally-aware iML tools (Heskes et al., 2020; Karimi et al., 2020; Wang et al., 2021), indicating that researchers in computational statistics are increasingly sensitive to the distinction between model- and system-level analyses. Genomic practitioners should avail themselves of both explanatory modes, but always make sure the selected tool matches the stated aim. Addressing this challenge is difficult at both a conceptual level, because the distinction between model- and system-level analyses may not be immediately obvious to practitioners, and at a technical level, because causal approaches can require careful covariate adjustments and data reweighting. The sooner these issues are addressed head on, the more fruitful the results will be.

### §5.2  Error Rate Control

Another open challenge in iML concerns bounding error and quantifying uncertainty. Bioinformaticians are no strangers to *p*-values, which are typically fixed at low levels to control false positive rates in GWAS (Panagiotou & Ioannidis, 2012), or else adjusted to control familywise error rates (Holm, 1979) or FDR (Benjamini & Hochberg, 1995) in other omic settings. Bayesians have their own set of inferential procedures for multiple testing scenarios (Gelman et al., 2012; Scott & Berger, 2010), although there is some notable convergence with frequentism on the subject of *q*-values (Storey, 2003). In any event, the error-statistical logic that guides testing in computational biology is largely absent from contemporary iML. This may be partially a result of cultural factors. As Breiman (2001b) observed some 20 years ago, there are two main cultures of statistical modeling – one focused on predicting outcomes, the other on inferring parameter values. Authors in contemporary iML, which grew almost exclusively out of the former camp, are generally less worried about error rates than their colleagues in the latter camp.

Several critics have pointed out that post-hoc methods do not generally provide standard errors for their estimates or goodness of fit measures for their approximations (Ribeiro et al., 2018; Wachter et al., 2018). Indeed, it would be difficult to do so without some nonparametric resampling procedure such as the bootstrap (Davison & Hinkley, 1997), which would add considerable computational burden as the number of samples and/or features grows. It is not clear that such methods are even applicable in these settings, however, given the instability of bootstrap estimators in high dimensions (Karoui & Purdom, 2018). This makes it impossible to reliably rank biomarkers or evaluate the reliability of experimental results. Knockoffs are a notable exception, given their focus on FDR control. However, the Candès et al. (2018) algorithm is not, strictly speaking, a post-hoc iML method, since it requires training an algorithm on an expanded $n \times 2p$ design matrix that includes both the original features and their knockoffs. This is just a preliminary step toward a final model, which contains only those variables that pass some predetermined FDR threshold. The modified method of Watson & Wright (2021) adapts knockoffs for post-hoc importance, but in so doing loses the finite sample error rate guarantees of the original adaptive thresholding procedure.

A handful of other iML methods make at least a nominal effort to quantify uncertainty (Gimenez & Zou, 2019; Ribeiro et al., 2018; Schwab & Karlen, 2019). Yet these examples are perhaps most notable for their scarcity. To gain more widespread acceptance in genomics – and the sciences more generally – iML algorithms will need to elevate rigorous testing procedures from an occasional novelty to a core requirement. Generic methods for doing so in high dimensions raise complex statistical challenges that remain unresolved at present.

### §5.3 Variable Granularity

A final challenge I will highlight concerns variable granularity. This is not a major issue in the low- or moderate-dimensional settings for which most iML tools are designed. But it quickly becomes important as covariates increase, especially when natural feature groupings are either known *a priori* or directly estimable from the data. For instance, it is well-established that genes do not operate in isolation, but rather work together in co-regulated pathways. Thus, even when a classifier uses gene-level RNA-seq data as input features, researchers may want to investigate the prognostic value of pathways to test or develop new hypotheses. In multi-omic models, where features typically represent a range of biological processes, each measured using different platforms, analysts may want to know not just which variables are most predictive overall, but which biomarkers are strongest within a given class. Interactions across subsystems may also be of particular interest.

Few methods in use today allow users to query a target model at varying degrees of resolution like this, but such flexibility would be a major asset in systems biology. Once again, some exceptions are worth noting. Sesia et al. (2020) introduce a knockoff method for localizing causal variants at different resolutions using well-established models of linkage disequilibrium. This amounts to a global post-hoc feature attribution method for whole genome sequencing data. The leave-out-covariates (LOCO) statistic (Lei et al., 2018; Rinaldo et al., 2019) can be used to quantify the global or local importance of arbitrary feature subsets, but only at the cost of extensive model refitting. Groupwise Shapley values have been formally described (Conitzer & Sandholm, 2004) but are not widely used in practice. Resolving the granularity problem will help iML tools scale better in high-dimensional settings, with major implications for genomics. The problem is complicated, however, by the fact that hierarchical information regarding biomolecular function is not always available. Automated methods for discovering such hierarchies are prone to error, while data-driven dimensionality reduction techniques – e.g., the latent embeddings learned by a deep neural network – can be difficult or impossible to interpret. Promising directions of research in this area include causal coarsening techniques (Beckers et al., 2019; Chalupka et al., 2017) and disentangled representation learning (Locatello et al., 2019; Schölkopf et al., 2021).

## §6 Conclusion

The pace of advances in genomics and ML can make it easy to forget that both disciplines are relatively young. The subfield of iML is even younger, with the vast majority of work published in just the last three to five years. The achievements to date at the intersection of these research programs are numerous and varied. Feature attributions and rule lists have already revealed novel insights in several genomic studies. Exemplary methods have not yet seen similar uptake, but that will likely change with better generative models. As datasets grow larger, computers become faster, and theoretical refinements continue to accumulate in statistics and biology, iML will become an increasingly integral part of the genomics toolkit. I have argued that better algorithmic explanations can serve researchers in at least three distinct ways: by auditing models for potential bias, validating performance before and throughout deployment, and revealing novel mechanisms for further exploration. I provided a simple typology for iML and reviewed several popular methodologies with a focus on genomic applications.

Despite considerable progress and rapidly expanding research interest, iML still faces a number of important conceptual and technical challenges. I highlighted three with particular significance for genomics: ambiguous targets, limited error rate control, and inflexible feature resolution. These obstacles can be addressed by iML solutions in a post-hoc or intrinsic manner, with model-agnostic or model-specific approaches, via global or local explanations. All types of iML require further development, especially as research in supervised learning and genomics continues to evolve. Ideally, iML would become integrated into standard research practice, part of hypothesis generation and testing as well as model training and deployment. As the examples from Sect. 4 illustrate, this vision is already on its way to becoming a reality.

The future of iML for genomics is bright. The last few years alone have seen a rapid proliferation of doctoral dissertations on the topic – e.g., Greenside (2018); Danaee (2019); Nikumbh (2019); Kavvas (2020); Ploenzke (2020); and Shrikumar (2020) – suggesting that early career academics in particular are being drawn to this highly interdisciplinary area of research. Existing work has been promising, though not without its challenges. As the field continues to gather more data, resources, and brainpower, there is every reason to believe the best is yet to come.

# References

Aas, K., Jullum, M., & Løland, A. (2019). Explaining individual predictions when features are dependent: More accurate approximations to Shapley values. *arXiv* preprint 1903.10464.

Adadi, A., & Berrada, M. (2018). Peeking Inside the Black-Box: A Survey on Explainable Artificial Intelligence (XAI). *IEEE Access*, *6*, 52138–52160.

Altmann, A., Toloşi, L., Sander, O., & Lengauer, T. (2010). Permutation importance: A corrected feature importance measure. *Bioinformatics*, *26*(10), 1340–1347.

Angelino, E., Larus-Stone, N., Alabi, D., Seltzer, M., & Rudin, C. (2018). Learning Certifiably Optimal Rule Lists for Categorical Data. *Journal of Machine Learning Research*, *18*(234), 1–78.

Anguita-Ruiz, A., Segura-Delgado, A., Alcalá, R., Aguilera, C. M., & Alcalá-Fdez, J. (2020). eXplainable Artificial Intelligence (XAI) for the identification of biologically relevant gene expression patterns in longitudinal human studies, insights from obesity research. *PLOS Computational Biology*, *16*(4), e1007792.

Azodi, C. B., Tang, J., & Shiu, S.-H. (2020). Opening the Black Box: Interpretable Machine Learning for Geneticists. *Trends in Genetics*, *36*(6), 442–455.

Bach, S., Binder, A., Montavon, G., Klauschen, F., Müller, K. R., & Samek, W. (2015). On pixel-wise explanations for non-linear classifier decisions by layer-wise relevance propagation. *PLoS ONE*, *10*(7), 1–46.

Barber, R. F., & Candès, E. J. (2015). Controlling the false discovery rate via knockoffs. *Ann. Statist.*, *43*(5), 2055–2085.

Bareinboim, E., & Pearl, J. (2016). Causal inference and the data-fusion problem. *Proceedings of the National Academy of Sciences*, *113*(27), 7345–7352.

Barredo Arrieta, A., Díaz-Rodríguez, N., Del Ser, J., Bennetot, A., Tabik, S., Barbado, A., … Herrera, F. (2020). Explainable Artificial Intelligence (XAI): Concepts, taxonomies, opportunities and challenges toward responsible AI. *Information Fusion*, *58*, 82–115.

Beckers, S., Eberhardt, F., & Halpern, J. Y. (2019). Approximate Causal Abstraction. *Proceedings of the International Conference on Uncertainty in Artificial Intelligence*, 210.

Benjamini, Y., & Hochberg, Y. (1995). Controlling the False Discovery Rate : A Practical and Powerful Approach to Multiple Testing. *Journal of the Royal Statistical Society: Series B (Statistical Methodology)*, *57*(1), 289–300.

Bhatt, U., Xiang, A., Sharma, S., Weller, A., Taly, A., Jia, Y., … Eckersley, P. (2020). Explainable Machine Learning in Deployment. *Proceedings of the 2020 Conference on Fairness, Accountability, and Transparency*, 648–657.

Breiman, L. (2001a). Random Forests. *Machine Learning*, *45*(1), 1–33.

Breiman, L. (2001b). Statistical Modeling: The Two Cultures. *Statist. Sci.*, *16*(3), 199–231.

Breiman, L., Friedman, J., Stone, C. J., & Olshen, R. A. (1984). *Classification and Regression Trees*. Boca Raton, FL: Taylor & Francis.

Candès, E., Fan, Y., Janson, L., & Lv, J. (2018). Panning for gold: 'model-X' knockoffs for high dimensional controlled variable selection. *Journal of the Royal Statistical Society: Series B (Statistical Methodology)*, *80*(3), 551–577.

Carrieri, A. P., Haiminen, N., Maudsley-Barton, S., Gardiner, L.-J., Murphy, B., Mayes, A. E., … Pyzer-Knapp, E. O. (2021). Explainable AI reveals changes in skin microbiome composition linked to phenotypic differences. *Scientific Reports*, *11*(1), 4565.

Caruana, R., Lou, Y., Gehrke, J., Koch, P., Sturm, M., & Elhadad, N. (2015). Intelligible Models for HealthCare. *Proceedings of the 21th ACM SIGKDD International Conference on Knowledge Discovery and Data Mining*, 1721–1730.

Carvalho, C. M., Polson, N. G., & Scott, J. G. (2010). The horseshoe estimator for sparse signals. *Biometrika*, *97*(2), 465–480.

Casalicchio, G., Molnar, C., & Bischl, B. (2019). Visualizing the Feature Importance for Black Box Models. In M. Berlingerio, F. Bonchi, T. Gärtner, N. Hurley, & G. Ifrim (Eds.), *Machine Learning and Knowledge Discovery in Databases* (pp. 655–670). Cham: Springer International Publishing.

Cascianelli, S., Molineris, I., Isella, C., Masseroli, M., & Medico, E. (2020). Machine learning for RNA sequencing-based intrinsic subtyping of breast cancer. *Scientific Reports*, *10*(1), 14071.

Chalupka, K., Eberhardt, F., & Perona, P. (2017). Causal feature learning: an overview. *Behaviormetrika*, *44*(1), 137–164.

Chen, C., Li, O., Tao, D., Barnett, A., Rudin, C., & Su, J. K. (2019). This Looks Like That: Deep Learning for Interpretable Image Recognition. *Advances in Neural Information Processing Systems 32* (pp. 8930–8941).

Chen, C., & Rudin, C. (2018). An Optimization Approach to Learning Falling Rule Lists. *Proceedings of the International Conference on Artificial Intelligence and Statistics* (pp. 604–612).

Chen, X., & Ishwaran, H. (2012). Random forests for genomic data analysis. *Genomics*, *99*(6), 323–329.

Chia, C., Sesia, M., Ho, C., Jeffrey, S., Dionne, J., Candès, E., & Howe, R. (2020). Interpretable classification of bacterial Raman spectra with knockoff wavelets. *arXiv* preprint 2006.04937.

Cliff, A., Romero, J., Kainer, D., Walker, A., Furches, A., & Jacobson, D. (2019). A High-Performance Computing Implementation of Iterative Random Forest for the Creation of Predictive Expression Networks. *Genes* , Vol. 10.

Conitzer, V., & Sandholm, T. (2004). Computing Shapley Values, Manipulating Value Division

Schemes, and Checking Core Membership in Multi-Issue Domains. *Proceedings of the 19th National Conference on Artifical Intelligence*, 219–225.

Danaee, P. (2019). *Interpretable machine learning: Applications in biology and genomics.* Doctoral dissertation, Oregon State University.

Das, A., & Rad, P. (2020). Opportunities and Challenges in Explainable Artificial Intelligence (XAI): A Survey. *arXiv* preprint, 2006.11371.

Davison, A. C., & Hinkley, D. V. (1997). *Bootstrap Methods and their Application*. Cambridge: Cambridge University Press.

Drouin, A., Letarte, G., Raymond, F., Marchand, M., Corbeil, J., & Laviolette, F. (2019). Interpretable genotype-to-phenotype classifiers with performance guarantees. *Scientific Reports*, *9*(1), 4071.

Edwards, L., & Veale, M. (2017). Slave to the algorithm? Why a "right to explanation" is probably not the remedy you are looking for. *Duke Law and Technology Review*, *16*(1), 18–84.

Egede, L. E., & Walker, R. J. (2020). Structural Racism, Social Risk Factors, and Covid-19 — A Dangerous Convergence for Black Americans. *New England Journal of Medicine*, *383*(12), e77.

Eraslan, G., Avsec, Ž., Gagneur, J., & Theis, F. J. (2019). Deep learning: new computational modelling techniques for genomics. *Nature Reviews Genetics*, *20*(7), 389–403.

Esteva, A., Kuprel, B., Novoa, R. A., Ko, J., Swetter, S. M., Blau, H. M., & Thrun, S. (2017). Dermatologist-level classification of skin cancer with deep neural networks. *Nature*, *542*(7639), 115–118.

Fisher, A., Rudin, C., & Dominici, F. (2019). All Models are Wrong, but Many are Useful: Learning a Variable's Importance by Studying an Entire Class of Prediction Models Simultaneously. *Journal of Machine Learning Research*, *20*(177), 1–81.

Friedman, J. H. (2001). Greedy Function Approximation: A Gradient Boosting Machine. *The Annals of Statistics*, *29*(5), 1189–1232.

Friedman, J. H., & Popescu, B. E. (2008). Predictive Learning via Rule Ensembles. *The Annals of Applied Statistics*, *2*(3), 916–954.

Garvin, M. R., T. Prates, E., Pavicic, M., Jones, P., Amos, B. K., Geiger, A., … Jacobson, D. (2020). Potentially adaptive SARS-CoV-2 mutations discovered with novel spatiotemporal and explainable AI models. *Genome Biology*, *21*(1), 304.

Gelman, A., Hill, J., & Yajima, M. (2012). Why We (Usually) Don't Have to Worry About Multiple Comparisons. *Journal of Research on Educational Effectiveness*, *5*(2), 189–211.

Gilpin, L. H., Bau, D., Yuan, B. Z., Bajwa, A., Specter, M., & Kagal, L. (2018). Explaining Explanations: An Overview of Interpretability of Machine Learning. *2018 IEEE 5th International Conference on Data Science and Advanced Analytics (DSAA)*, 80–89.

Gimenez, J. R., & Zou, J. (2019). Discovering Conditionally Salient Features with Statistical Guarantees. *Proceedings of the 36th International Conference on Machine Learning* (pp. 2290–2298).

Greenside, P. (2018). *Interpretable machine learning methods for regulatory and disease genomics*. Doctoral dissertation, Stanford University.

Guidotti, R., Monreale, A., Ruggieri, S., Pedreschi, D., Turini, F., & Giannotti, F. (2018). Local Rule-Based Explanations of Black Box Decision Systems. *arXiv* preprint, 1805.10820.

Guidotti, R., Monreale, A., Ruggieri, S., Turini, F., Giannotti, F., & Pedreschi, D. (2018). A Survey of Methods for Explaining Black Box Models. *ACM Computing Surveys*, *51*(5), 1–42.

Hastie, T., Tibshirani, R., & Friedman, J. (2009). *The Elements of Statistical Learning: Data Mining, Inference, and Prediction*. New York: Springer.

Heinze-Deml, C., Peters, J., & Meinshausen, N. (2018). Invariant Causal Prediction for Nonlinear Models. *Journal of Causal Inference*, *6*(2).

Heskes, T., Sijben, E., Bucur, I. G., & Claassen, T. (2020). Causal Shapley Values: Exploiting Causal Knowledge to Explain Individual Predictions of Complex Models. *Advances in Neural Information Processing Systems*.

Holm, S. (1979). A Simple Sequentially Rejective Multiple Test Procedure. *Scandinavian Journal of Statistics*, *6*(2), 65–70.

Holzinger, A., Langs, G., Denk, H., Zatloukal, K., & Müller, H. (2019). Causability and explainability of artificial intelligence in medicine. *WIREs Data Mining and Knowledge Discovery*, *9*(4), e1312.

Hu, X., Rudin, C., & Seltzer, M. (2019). Optimal Sparse Decision Trees. *Advances in Neural Information Processing Systems 32* (pp. 7267–7275).

Hubbard, A. E., Kennedy, C. J., & van der Laan, M. J. (2018). Data-adaptive target parameters. In M. J. van der Laan & S. Rose (Eds.), *Targeted Learning in Data Science* (pp. 125–142). New York: Springer.

Hyafil, L., & Rivest, R. L. (1976). Constructing optimal binary decision trees is NP-complete. *Information Processing Letters*, *5*(1), 15–17.

Imbens, G. W., & Rubin, D. B. (2015). *Causal Inference for Statistics, Social, and Biomedical Sciences: An Introduction*. Cambridge: Cambridge University Press.

Karimi, A.-H., Barthe, G., Schölkopf, B., & Valera, I. (2020). A survey of algorithmic recourse: definitions, formulations, solutions, and prospects. *arXiv* preprint, 2010.04050.

Karimi, A.-H., von Kügelgen, J., Schölkopf, B., & Valera, I. (2020). Algorithmic recourse under imperfect causal knowledge: a probabilistic approach. *Advances in Neural Information Processing Systems*.

Karoui, N. El, & Purdom, E. (2018). Can We Trust the Bootstrap in High-dimensions? The Case of Linear Models. *Journal of Machine Learning Research*, *19*(5), 1–66.

Kavvas, E. (2020). *Biologically-interpretable machine learning for microbial genomics*. Doctoral dissertation, UC San Diego.

Kavvas, E. S., Yang, L., Monk, J. M., Heckmann, D., & Palsson, B. O. (2020). A biochemically-interpretable machine learning classifier for microbial GWAS. *Nature Communications*, *11*(1), 2580.

Lage, I., Chen, E., He, J., Narayanan, M., Gershman, S., Kim, B., & Doshi-Velez, F. (2018). An Evaluation of the Human-Interpretability of Explanation. *Conference on Neural Information Processing Systems (NeurIPS) Workshop on Correcting and Critiquing Trends in Machine Learning*.

Lakkaraju, H., Kamar, E., Caruana, R., & Leskovec, J. (2019). Faithful and Customizable Explanations of Black Box Models. *Proceedings of the 2019 AAAI/ACM Conference on AI, Ethics, and Society*, 131–138.

Law, C. W., Chen, Y., Shi, W., & Smyth, G. K. (2014). voom: precision weights unlock linear model analysis tools for RNA-seq read counts. *Genome Biology*, *15*(2), R29.

Lei, J., G'Sell, M., Rinaldo, A., Tibshirani, R. J., & Wasserman, L. (2018). Distribution-Free Predictive Inference for Regression. *Journal of the American Statistical Association*, *113*(523), 1094–1111.

Letham, B., Rudin, C., McCormick, T. H., & Madigan, D. (2015). Interpretable classifiers using rules and Bayesian analysis: Building a better stroke prediction model. *Ann. Appl. Stat.*, *9*(3), 1350–1371.

Levy, J. J., Titus, A. J., Petersen, C. L., Chen, Y., Salas, L. A., & Christensen, B. C. (2020). MethylNet: an automated and modular deep learning approach for DNA methylation analysis. *BMC Bioinformatics*, *21*(1), 108.

Li, S., Sesia, M., Romano, Y., Candès, E., & Sabatti, C. (2021). Searching for consistent associations with a multi-environment knockoff filter. *arXiv* preprint, 2106.04118.

Linardatos, P., Papastefanopoulos, V., & Kotsiantis, S. (2021). Explainable AI: A Review of Machine Learning Interpretability Methods. *Entropy*, Vol. 23.

Locatello, F., Bauer, S., Lucic, M., Raetsch, G., Gelly, S., Schölkopf, B., & Bachem, O. (2019). Challenging Common Assumptions in the Unsupervised Learning of Disentangled Representations. *Proceedings of the International Conference on Machine Learning*.

Love, M. I., Huber, W., & Anders, S. (2014). Moderated estimation of fold change and dispersion for RNA-seq data with DESeq2. *Genome Biology*, *15*(12), 550.

Lundberg, S. M., Erion, G., Chen, H., DeGrave, A., Prutkin, J. M., Nair, B., … Lee, S.-I. (2020). From local explanations to global understanding with explainable AI for trees. *Nature Machine Intelligence*, *2*(1), 56–67.

Lundberg, S. M., & Lee, S.-I. (2017). A Unified Approach to Interpreting Model Predictions.

*Advances in Neural Information Processing Systems 30* (pp. 4765–4774).

Martin, A. R., Kanai, M., Kamatani, Y., Okada, Y., Neale, B. M., & Daly, M. J. (2019). Clinical use of current polygenic risk scores may exacerbate health disparities. *Nature Genetics*, *51*(4), 584–591.

Mazurowski, M. A., Buda, M., Saha, A., & Bashir, M. R. (2019). Deep learning in radiology: An overview of the concepts and a survey of the state of the art with focus on MRI. *Journal of Magnetic Resonance Imaging*, *49*(4), 939–954.

McKinney, S. M., Sieniek, M., Godbole, V., Godwin, J., Antropova, N., Ashrafian, H., … Shetty, S. (2020). International evaluation of an AI system for breast cancer screening. *Nature*, *577*(7788), 89–94.

Merrick, L., & Taly, A. (2020). The Explanation Game: Explaining Machine Learning Models Using Shapley Values. *Machine Learning and Knowledge Extraction - 4th International Cross-Domain Conference (CD-MAKE)* (pp. 17–38).

Milioli, H. H., Tishchenko, I., Riveros, C., Berretta, R., & Moscato, P. (2017). Basal-like breast cancer: molecular profiles, clinical features and survival outcomes. *BMC Medical Genomics*, *10*(1), 19.

Mohseni, S., Zarei, N., & Ragan, E. D. (2020). A Multidisciplinary Survey and Framework for Design and Evaluation of Explainable AI Systems. *arXiv* preprint, 1811.11839.

Molnar, C. (2020). *Interpretable Machine Learning: A Guide for Making Black Box Models Interpretable*. Münich: Christoph Molnar.

Montavon, G., Lapuschkin, S., Binder, A., Samek, W., & Müller, K. R. (2017). Explaining nonlinear classification decisions with deep Taylor decomposition. *Pattern Recognition*, *65*(May 2016), 211–222.

Mueller, S. T., Hoffman, R. R., Clancey, W., Emrey, A., & Klein, G. (2019). Explanation in Human-AI Systems: A Literature Meta-Review, Synopsis of Key Ideas and Publications, and Bibliography for Explainable AI. *arXiv* preprint, 1902.01876.

Murdoch, W. J., Singh, C., Kumbier, K., Abbasi-Asl, R., & Yu, B. (2019). Definitions, methods, and applications in interpretable machine learning. *Proceedings of the National Academy of Sciences*, *116*(44), 22071–22080.

Nalenz, M., & Villani, M. (2018). Tree ensembles with rule structured horseshoe regularization. *Ann. Appl. Stat.*, *12*(4), 2379–2408.

Nembrini, S., König, I. R., & Wright, M. N. (2018). The revival of the Gini importance? *Bioinformatics*, *34*(21), 3711–3718.

Nicholls, H. L., John, C. R., Watson, D. S., Munroe, P. B., Barnes, M. R., & Cabrera, C. P. (2020). Reaching the End-Game for GWAS: Machine Learning Approaches for the Prioritization of Complex Disease Loci. *Frontiers in Genetics*, *11*, 350.

Nikumbh, S. (2019). *Interpretable machine learning methods for prediction and analysis of*

*genome regulation in 3D*. Doctoral dissertation, Saarland University.

Obermeyer, Z., Powers, B., Vogeli, C., & Mullainathan, S. (2019). Dissecting racial bias in an algorithm used to manage the health of populations. *Science*, *366*(6464), 447–453.

Panagiotou, O. A., & Ioannidis, J. P. A. (2012). What should the genome-wide significance threshold be? Empirical replication of borderline genetic associations. *International Journal of Epidemiology*, *41*(1), 273–286.

Pearl, J. (2000). *Causality: Models, Reasoning, and Inference*. New York: Cambridge University Press.

Pearl, J., & Bareinboim, E. (2014). External Validity: From Do-Calculus to Transportability Across Populations. *Statist. Sci.*, *29*(4), 579–595.

Peters, J., Bühlmann, P., & Meinshausen, N. (2016). Causal inference by using invariant prediction: identification and confidence intervals. *Journal of the Royal Statistical Society: Series B (Statistical Methodology)*, *78*(5), 947–1012.

Peters, J., Janzing, D., & Schölkopf, B. (2017). *The Elements of Causal Inference: Foundations and Learning Algorithms*. Cambridge, MA: The MIT Press.

Pfister, N., Bühlmann, P., & Peters, J. (2019). Invariant Causal Prediction for Sequential Data. *Journal of the American Statistical Association*, *114*(527), 1264–1276.

Ploenzke, M. (2020). *Interpretable machine learning methods with applications in genomics*. Doctoral dissertation, Harvard University.

Ribeiro, M. T., Singh, S., & Guestrin, C. (2016). "Why Should I Trust You?": Explaining the Predictions of Any Classifier. *Proceedings of the 22nd ACM SIGKDD International Conference on Knowledge Discovery and Data Mining*, 1135–1144.

Ribeiro, M. T., Singh, S., & Guestrin, C. (2018). Anchors: High-Precision Model-Agnostic Explanations. *AAAI*, 1527–1535.

Rinaldo, A., Wasserman, L., & G'Sell, M. (2019). Bootstrapping and sample splitting for high-dimensional, assumption-lean inference. *Ann. Statist.*, *47*(6), 3438–3469.

Rudin, C. (2019). Stop explaining black box machine learning models for high stakes decisions and use interpretable models instead. *Nature Machine Intelligence*, *1*(5), 206–215.

Sarkar, J. P., Saha, I., Sarkar, A., & Maulik, U. (2021). Machine learning integrated ensemble of feature selection methods followed by survival analysis for predicting breast cancer subtype specific miRNA biomarkers. *Computers in Biology and Medicine*, *131*, 104244.

Schölkopf, B, Locatello, F., Bauer, S., Ke, N. R., Kalchbrenner, N., Goyal, A., & Bengio, Y. (2021). Toward Causal Representation Learning. *Proceedings of the IEEE*, *109*(5), 612–634.

Schölkopf, Bernhard, Tsuda, K., & Vert, J.-P. (Eds.). (2004). *Kernel Methods in Computational Biology*. Cambridge, MA: The MIT Press.

Schwab, P., & Karlen, W. (2019). CXPlain: Causal Explanations for Model Interpretation

under Uncertainty. *Advances in Neural Information Processing Systems 32* (pp. 10220–10230).

Scott, J. G., & Berger, J. O. (2010). Bayes and empirical-Bayes multiplicity adjustment in the variable-selection problem. *The Annals of Statistics*, *38*(5), 2587–2619.

Selbst, A., & Powles, J. (2017). Meaningful information and the right to explanation. *International Data Privacy Law*, *7*(4), 233–242.

Shah, R. D., & Meinshausen, N. (2014). Random Intersection Trees. *Journal of Machine Learning Research*, *15*(20), 629–654.

Shapley, L. (1953). A Value for n-Person Games. In *Contributions to the Theory of Games* (pp. 307–317).

Shrikumar, A. (2020). *Interpretable machine learning for scientific discovery in regulatory genomics*. Doctoral dissertation, Stanford University.

Shrikumar, A., Greenside, P., & Kundaje, A. (2017). Learning Important Features Through Propagating Activation Differences. *Proceedings of the International Conference on Machine Learning*.

Smyth, G. K. (2004). Linear models and empirical Bayes methods for assessing differential expression in microarray experiments. *Stat Appl Genet Mol Biol*, *3*.

Sonnenburg, S., Zien, A., Philips, P., & Rätsch, G. (2008). POIMs: positional oligomer importance matrices—understanding support vector machine-based signal detectors. *Bioinformatics*, *24*(13), i6–i14.

Sørlie, T., Tibshirani, R., Parker, J., Hastie, T., Marron, J. S., Nobel, A., … Botstein, D. (2003). Repeated observation of breast tumor subtypes in independent gene expression data sets. *Proceedings of the National Academy of Sciences*, *100*(14), 8418 LP – 8423.

Stokes, J. M., Yang, K., Swanson, K., Jin, W., Cubillos-Ruiz, A., Donghia, N. M., … Collins, J. J. (2020). A Deep Learning Approach to Antibiotic Discovery. *Cell*, *180*(4), 688-702.e13.

Storey, J. D. (2003). The positive false discovery rate: a Bayesian interpretation and the q - value. *Ann. Statist.*, *31*(6), 2013–2035.

Strobl, C., Boulesteix, A.-L., Kneib, T., Augustin, T., & Zeileis, A. (2008). Conditional Variable Importance for Random Forests. *BMC Bioinformatics*, *9*(1), 307.

Sundararajan, M., & Najmi, A. (2019). The many Shapley values for model explanation. *Proceedings of the ACM Conference*. New York: ACM.

Sundararajan, M., Taly, A., & Yan, Q. (2017). Axiomatic Attribution for Deep Networks. *Proceedings of the International Conference on Machine Learning*.

Talukder, A., Barham, C., Li, X., & Hu, H. (2020). Interpretation of deep learning in genomics and epigenomics. *Briefings in Bioinformatics*.

Tasaki, S., Gaiteri, C., Mostafavi, S., & Wang, Y. (2020). Deep learning decodes the principles of differential gene expression. *Nature Machine Intelligence*, *2*(7), 376–386.

Tibshirani, R. (1996). Regression Shrinkage and Selection via the Lasso. *Journal of the Royal Statistical Society. Series B (Methodological)*, *58*(1), 267–288.

Tideman, L. E. M., Migas, L. G., Djambazova, K. V, Patterson, N. H., Caprioli, R. M., Spraggins, J. M., & Van de Plas, R. (2020). Automated Biomarker Candidate Discovery in Imaging Mass Spectrometry Data Through Spatially Localized Shapley Additive Explanations. *BioRxiv*.

Topol, E. J. (2019). High-performance medicine: the convergence of human and artificial intelligence. *Nature Medicine*, *25*(1), 44–56.

Turner, N. C., & Reis-Filho, J. S. (2006). Basal-like breast cancer and the BRCA1 phenotype. *Oncogene*, *25*, 5846.

Vapnik, V. (1995). *The Nature of Statistical Learning Theory*. New York: Springer-Verlag.

Vidovic, M. M.-C., Görnitz, N., Müller, K.-R., Rätsch, G., & Kloft, M. (2015). SVM2Motif--Reconstructing Overlapping DNA Sequence Motifs by Mimicking an SVM Predictor. *PloS One*, *10*(12), e0144782–e0144782.

Vidovic, M. M.-C., Kloft, M., Müller, K.-R., & Görnitz, N. (2017). ML2Motif-Reliable extraction of discriminative sequence motifs from learning machines. *PloS One*, *12*(3), e0174392–e0174392.

Vilone, G., & Longo, L. (2020). Explainable Artificial Intelligence: a Systematic Review. *arXiv* preprint, 2006.00093.

Wachter, S., Mittelstadt, B., & Floridi, L. (2017). Why a right to explanation of automated decision-making does not exist in the general data protection regulation. *International Data Privacy Law*, *7*(2), 76–99.

Wachter, S., Mittelstadt, B., & Russell, C. (2018). Counterfactual Explanations without Opening the Black Box: Automated Decisions and the GDPR. *Harvard Journal of Law and Technology*, *31*(2), 841–887.

Waldmann, P., Mészáros, G., Gredler, B., Fürst, C., & Sölkner, J. (2013). Evaluation of the lasso and the elastic net in genome-wide association studies. *Frontiers in Genetics*, *4*, 270.

Watson, D., Krutzinna, J., Bruce, I. N., Griffiths, C. E. M., McInnes, I. B., Barnes, M. R., & Floridi, L. (2019). Clinical applications of machine learning algorithms: beyond the black box. *BMJ*, *364*, 446-448.

Watson, D. S., & Floridi, L. (2020). The explanation game: a formal framework for interpretable machine learning. *Synthese*.

Williamson, B. D., Gilbert, P. B., Carone, M., & Simon, N. (2020). Nonparametric variable importance assessment using machine learning techniques. *Biometrics*.

Woodward, J. (2019). Scientific Explanation. In E. N. Zalta (Ed.), *The Stanford Encyclopedia of Philosophy* (Winter 201). Metaphysics Research Lab, Stanford University.

Xie, Y. R., Castro, D. C., Bell, S. E., Rubakhin, S. S., & Sweedler, J. V. (2020). Single-Cell Classification Using Mass Spectrometry through Interpretable Machine Learning. *Analytical Chemistry*, *92*(13), 9338–9347.

Xu, G., Duong, T. D., Li, Q., Liu, S., & Wang, X. (2020). Causality Learning: A New Perspective for Interpretable Machine Learning. *arXiv* preprint, 2006.16789.

Yang, H., Rudin, C., & Seltzer, M. (2017). Scalable Bayesian Rule Lists. *Proceedings of the International Conference on Machine Learning*.

Yap, M., Johnston, R. L., Foley, H., MacDonald, S., Kondrashova, O., Tran, K. A., … Waddell, N. (2021). Verifying explainability of a deep learning tissue classifier trained on RNA-seq data. *Scientific Reports*, *11*(1), 2641.

Zhao, Q., & Hastie, T. (2019). Causal Interpretations of Black-Box Models. *Journal of Business & Economic Statistics*, 1–10.

Zou, H., & Hastie, T. (2005). Regularization and Variable Selection via the Elastic Net. *Journal of the Royal Statistical Society. Series B (Statistical Methodology)*, *67*(2), 301–320.